\documentclass[seceq]{ptptex}

\usepackage{graphicx}
\title{
Analytical Expression for Low-dimensional Resonance Island \\
in 4-dimensional Symplectic Map
}

\author{%       %Use \scshape  for the family name
Shin-itiro \textsc{Goto}%
}

\inst{%         %Affiliation, neglected when [addenda] or [errata]
Department of Applied Mathematics and Physics, 
Kyoto University, Kyoto 606-8501, Japan
}

\abst{
We study a $2$- and a $4$-dimensional nearly integrable symplectic maps using a singular perturbation method.
Resonance island structures in the $2$- and $4$- dimensional maps are 
successfully obtained. Those perturbative results are numerically confirmed.   
}

\begin{document}
%%%%%%%%%%%%%%%%%%%%%%%%%%%%%%%%%%%%%%%%%%
\newcommand{\beq}{\begin{equation}}
\newcommand{\beqa}{\begin{eqnarray}}
\newcommand{\eeq}{\end{equation}}
\newcommand{\eeqa}{\end{eqnarray}}
\newcommand{\non}{\nonumber}
\newcommand{\lb}{\label}
\newcommand{\fr}[1]{(\ref{#1})}
\newcommand{\ve}{{\varepsilon}}
%%%%%%%%%%%%%%%%%%%%%%%%%%%%%%%%%%%%%%%%%%
\maketitle
%%%%%%%%%%%%%%%%%%%%%%%%%%%%%%%%%%%%%%%%%%5
%\section{Section Title}
%Start your paper from here.
%%%%%%%%%%%%%%%%%%%%%%%%%%%%%%%%%%%%%%%%%%5
\section{Introduction}
%%%%%%%%%%%%%%%%%%%%%%%%%%%%%%%%%%%%%%%%%%5

Chaotic Hamiltonian systems with many degrees of freedom 
are not only of interest in view of the foundation of dynamical system, 
but also have a lot of applications.
Classical Hamiltonian mechanics 
has been applied to various physical problems, 
such as chemical reactions, plasma physics and solar systems\cite{LL92}. 
Although it is important to study such systems, 
we cannot understand the global flow generated by a Hamiltonian 
in the phase space due to the high-dimensionality. 
For a low-dimensional system, the phase space is visualized and then 
one overinterprets high-dimensional Hamiltonian flows beyond ones 
in a $2$-dimensional phase space, but such imaginations are not guaranteed.

On the other hand, 
the origin of the global Hamiltonian chaos is an overlap of two resonance 
islands in the phase space\cite{LL92}.
Then a fundamental problem for the study of Hamiltonian chaos
is to analyze a resonance structure. To tackle this problem for 
systems with many degrees of freedom, we take a singular perturbation method.
Using a systematic singular perturbation method, we can 
show the functional form of an island structure 
even for high-dimensional systems without information 
on the geometry of the phase space. 

In this article, we take a renormalization method which is one of 
singular perturbation methods.
All sorts of renormalization methods remove secular or divergent terms 
from a naive perturbation series by renormalized integration constants 
of the non-perturbation solution, and each 
prescription of the method does not depend on the detail 
of a given system\cite{CGO,Kun,GMN99,Oon00}.
One of the reformulated renormalization method which we take here is 
easily applied to non-chaotic systems and also 
chaotic maps\cite{GMN99,RGdiff,Gotos}.
Using this reformulated renormalization method, we tackle the standard map 
defined in $2$-dimensional phase and the Froeschl\'e map defined in 
$4$-dimensional phase space. 
Both maps are chaotic and are known for the standard models to 
study Hamiltonian chaos. 

%%%%%%%%%%%%%%%%%%%%%%%%%%%%%%%%%%%%%%%%%%5
\section{The Standard map}
\lb{sec:standard-map}
%%%%%%%%%%%%%%%%%%%%%%%%%%%%%%%%%%%%%%%%%%5

To begin with,
we study a resonance structure in the standard map \cite{LL92}
$(x^n,y^n)\mapsto(x^{n+1},y^{n+1})$
\beqa
x^{n+1}&=&x^n+y^{n+1},\quad (\mbox{mod} 2\pi)\non\\
y^{n+1}&=&y^n +\ve \sin(x^n).\quad (\mbox{mod} 2\pi) \non
\eeqa
Here $x^n$ and $y^n$ are the canonical variables, 
$n(\in\mathbb{Z})$ denotes the discrete time and $\ve (0<\ve\ll 1)$ is 
the small parameter, and this map is symplectic 
($dx^{n+1}\wedge dy^{n+1}=dx^n\wedge dy^n$).
Eliminating $y$ variable, we have
\beq
Lx^n=\ve\sin(x^n),\quad (\mbox{mod} 2\pi)
\lb{eqn:stdmp}
\eeq
where $Lx^n:= x^{n+1}-2x^n+x^{n-1}.$
A naive perturbation solution shows where a resonance island appears, and 
the structure can be obtained using our renormalization method. In our case,
the naive perturbation series expansion 
$x^n=x^{(0)n}+\ve x^{(1)n}+\ve^2x^{(2)n}+{\cal O}(\ve^3)$ 
is obtained by solving
$Lx^{(0)n}=0,Lx^{(1)n}=\sin(x^{(0)n}),Lx^{(2)n}=x^{(1)n}\cos(x^{(0)n}),\cdots$,
where (mod $2\pi$) operation is taken into account in each equation. 
The non-perturbative solution is given by $x^{(0)n}=a+nP$,
where $a$ and $P$ are the integration constants. 
Here $P$ is the value of $y^{(0)n}$, due to the relation 
$y^{(0)n+1}:=x^{(0)n+1}-x^{(0)n}=P$.  
Then the values of $P$ and $a$ are in $[0,1)$. 
The naive perturbation solution up to ${\cal O}(\ve^2)$ 
is classified by the value of $P$ as follows
\beqa
x^n&=&a+nP
+\ve\frac{-\sin(a+nP)}{4\sin^2(P/2)}
+\ve^2\frac{\sin(2(a+nP))}{32\sin^2(P)\sin^2(P/2)}\non\\
&&+{\cal O}(\ve^3),(\mbox{mod} 2\pi)
,\quad (P\neq 0,\pi)\non\\
x^n&=&a+\ve\frac{\sin(a)}{2}n^2
+\ve^2\frac{\sin(2a)}{48}(n^4-n^2)+{\cal O}(\ve^3)
,(\mbox{mod} 2\pi)\quad (P=0)\non\\
x^n&=&a+\pi n
+\ve\frac{-\sin(a+\pi n)}{4}
+\ve^2\frac{-\sin(2a)}{16}n^2\non\\
&&+{\cal O}(\ve^3),(\mbox{mod} 2\pi)\quad (P=\pi)
\lb{eqn:stdmp-naive}
\eeqa
Here we concentrate on the case of $P=\pi$ as an example. 
In the series \fr{eqn:stdmp-naive}, the secular behavior $(\propto \ve^2n^2)$ 
appears. 
The renormalization method removes the secular terms in a systematic way. 
First we define the renormalized variable $a^n$ as follows,
\beq
a^n:=a+\ve^2  \frac{-\sin(2a)}{16}n^2.
\lb{eqn:stdmp-def-an}
\eeq
This definition \fr{eqn:stdmp-def-an} should include all secular term up to 
${\cal O}(\ve^2)$ in \fr{eqn:stdmp-naive} and the 
integration constant, $a$, coming from the non-perturbative problem.
The way to define a renormalized variable is quite similar to one in 
the case of a differential equation\cite{GMN99}.
The renormalization equation is the map satisfying $a^n$ perturbatively, 
and we impose here 
that the renormalization map 
which we would like to obtain
is symplectic and autonomous. This is because 
the original map \fr{eqn:stdmp} has these two properties. 
From the definition of the renormalized variable \fr{eqn:stdmp-def-an}, 
we have 
$$
La^n=L\bigg(a+\ve^2\frac{-\sin(2a)}{16}n^2\bigg)=-\ve^2\frac{\sin(2a)}{8}.
$$ 
To obtain the map containing $a^n$ only,
we substitute $a=a^n+{\cal O}(\ve)$ coming 
from the definition \fr{eqn:stdmp-def-an} into the relation 
$La^n=-\ve^2\sin(2a)/8$. Then we have the following renormalization map,
\beq
La^n=-\ve^2\frac{\sin(2a^n)}{8}.
\lb{eqn:stdmp-rgmap}
\eeq
When we introduce the new variable $b^n$ defined 
as $b^{n+1}:=a^{n+1}-a^n$, it turns out that this 
renormalization map \fr{eqn:stdmp-rgmap} is automatically symplectic 
($da^{n+1}\wedge db^{n+1}=da^n\wedge db^n$). 
It is noted here that the reduced map \fr{eqn:stdmp-rgmap} is also 
obtained using another singular perturbation method, and 
the similarity of the two maps \fr{eqn:stdmp} and \fr{eqn:stdmp-rgmap} 
gives a self-similarity in the phase space \cite{BR83}.  

From Eqs. \fr{eqn:stdmp-naive} and \fr{eqn:stdmp-def-an}, 
the relations between 
the renormalized variable $a^n$ and the original variables $(x^n,y^n)$ 
are obtained as  
\beqa
x^n&=&a^n+\pi n -\frac{\ve}{4}\sin(a^n+\pi n)+{\cal O}(\ve^3),~(\mbox{mod} 2\pi)\non\\
y^n&=&(a^n-a^{n-1})+\pi -\frac{\ve}{2}\sin(a^n+\pi n)
+{\cal O}(\ve^3).\quad (\mbox{mod} 2\pi)\non
\eeqa
The unstable manifold of the renormalization map reproduces the resonance 
island located near $y=\pi$ in the original system \fr{eqn:stdmp}. 
Fig. \ref{fig:stdmp} shows that the phase space reconstructed 
using our renormalization method is close to 
one obtained by the original map even with $\ve=0.98$. 
Intuitively, this perturbative 
analysis is valid only for the case $0<\ve\ll 1$, however,
Fig.1 shows that the validity of our renormalization map seems to be wider 
than $0<\ve\ll 1$ practically. 
To determine the actual validity of this analysis, we  
need a rigorous mathematical estimate.
At any rate, 
it turns out that our renormalization method is also 
useful even for chaotic maps,
and that our analysis could not be restricted in $2$-dimensional maps.
%%%%%%%%%%%%%%%
\begin{figure}
\includegraphics[width=7cm]
%{/home/sgoto/liouville/island-str/stdmp-resonant-rg-paper1.eps}
{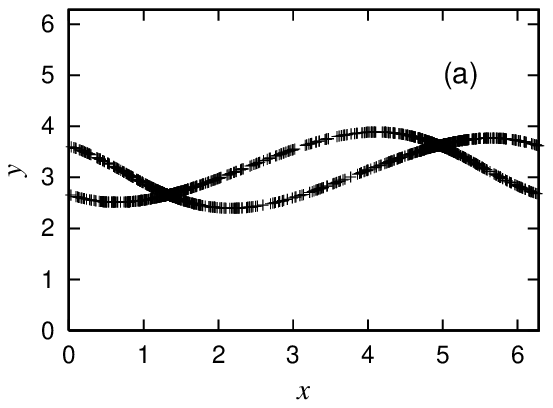}
\includegraphics[width=7cm]
%{/home/sgoto/liouville/island-str/stdmp-resonant-org-paper1.eps}
{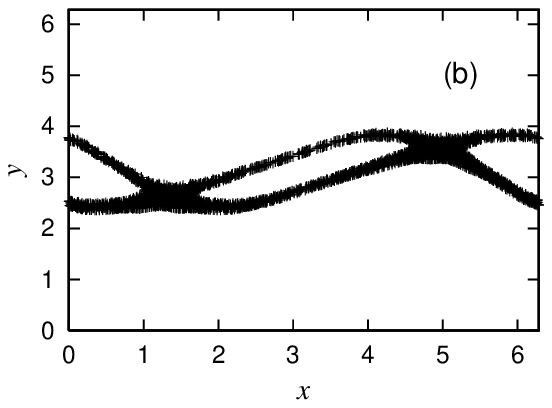}
\caption{
% THIS IS FOR PR 
%The reconstructed resonance island near $y=\pi$ using 
%the renormalization method, 
%the inset shows the one obtained by the original map 
%\fr{eqn:stdmp} . For both figures,  the value of $K$ is $0.98$, and 
%the number of the initial condition is $1$.
The resonance island near $y=\pi$.
For both figures the value of $\ve$ is $0.98$, and 
the number of the initial condition is $1$. 
(a): using the renormalization method, 
(b): obtained by the original map 
\fr{eqn:stdmp}.
}
\lb{fig:stdmp}  
\end{figure}
%%%%%%%%%%%%

%%%%%%%%%%%%%%%%%%%%%%%%%%%%%%%%%%%%%%%%%%5
\section{The Froeschl\'e map}
%%%%%%%%%%%%%%%%%%%%%%%%%%%%%%%%%%%%%%%%%%5

Next, we would like to study a $4$-dimensional symplectic map. 
It is expected that a 
similar analysis gives a functional form of a resonant island 
even in a high-dimensional map, and it will be confirmed later in this section.
The map which we study here is the Froeschl\'e map\cite{Froe71},    
$(x_1^n,x_2^n,y_1^n,y_2^n)\mapsto (x_1^{n+1},x_2^{n+1},y_1^{n+1},y_2^{n+1})$, 
$(\sum_{j=1}^2 dx_j^{n+1}\wedge dy_j^{n+1}=\sum_{j=1}^2 dx_j^n\wedge dy_j^n)$ 
\beqa
y_1^{n+1}&=&y_1^n+\frac{\ve A_1}{2\pi}\sin(2\pi x_1^n)+\frac{\ve C}{2\pi}\sin(2\pi(x_1^n+x_2^n)),\quad(\mbox{mod} 1)\non\\
y_2^{n+1}&=&y_2^n+\frac{\ve A_2}{2\pi}\sin(2\pi x_2^n)+
\frac{\ve C}{2\pi}\sin(2\pi(x_1^n+x_2^n)),\quad (\mbox{mod} 1)\non\\
x_1^{n+1}&=&x_1^n+y_1^{n+1},\quad (\mbox{mod} 1)\non\\
x_2^{n+1}&=&x_2^n+y_2^{n+1}.\quad (\mbox{mod} 1)\non
\eeqa
Here $n(\in\mathbb{Z})$ denotes the discrete time, 
$x_1^n,x_2^n$ and $y_1^n,y_2^n$ are the canonical variables,
$A_1,A_2,C$ are the parameters and $\ve (0<|\ve|\ll 1)$  
is the perturbation parameter. 
The Froeschl\'e map\cite{Froe71} has numerically been studied for 
the Arnold diffusion\cite{LL92} as one of canonical $4$-dimensional maps 
in several papers.\cite{FroeStudy}
Eliminating $y_1^n$ and $y_2^n$ variables, we have 
\beqa
Lx_j^n&=&\frac{\ve A_j}{2\pi}\sin(2\pi x_j^n)
+\frac{\ve C}{2\pi}\sin(2\pi(x_1^n+x_2^n)),(\mbox{mod} 1),\quad (j=1,2)\non
\eeqa
where $Lx_j^n:=x_j^{n+1}-2x_j^n+x_j^{n-1}$, $(j=1,2)$.
The naive perturbation solutions 
$x_j^n=x_j^{(0)n}+\ve x_j^{(1)n}+{\cal O}(\ve^2), (j=1,2)$ 
are obtained by solving the following equations,
\beqa
Lx_j^{(0)n}&=&0,(\mbox{mod}1)\quad (j=1,2)\non\\
Lx_j^{(1)n}&=&\frac{A_j}{2\pi}\sin(2\pi x_j^{(0)n})
+\frac{C}{2\pi}\sin(2\pi(x_1^{(0)n}+x_2^{(0)n})).
(\mbox{mod}1)~(j=1,2)\non
\eeqa
The non-perturbative solutions are $x_j^{(0)}=a_j+nP_j$, $(j=1,2)$. Here $a_j$
and $P_j$ are the integration constants, and the values of them are in $[0,1)$.
The naive perturbation solutions are 
classified by the values of $P_1$ and $P_2$, 
the solutions up to ${\cal O}(\ve)$ are 
\beqa
x_j^{n}&=& a_j+n P_j +
\ve\bigg(
\frac{A_j}{4\pi}\frac{\sin(2\pi (a_j+nP_j))}{\cos(2\pi P_j)-1}
+\frac{C}{4\pi}\frac{\sin(2\pi(a_1+nP_1)
+2\pi(a_2+nP_2)}{\cos(2\pi(P_1+P_2))-1}
\bigg)\non\\
&&+{\cal O}(\ve^2),(\mbox{mod} 1), 
(P_1\neq 0,P_2\neq 0, P_1+P_2\neq 0)
\lb{eqn:froe-naive-nr}\\
x_j^{n}&=& a_j+nP_j+
\ve\bigg(
\frac{A_j}{4\pi}\frac{\sin(2\pi (a_j+nP_j))}{\cos(2\pi P_j)-1}
+\frac{ C}{4\pi}n^2\sin(2\pi (a_1+a_2))\bigg)\non\\
&&+{\cal O}(\ve^2),(\mbox{mod} 1)\quad(P_1\neq 0,P_2\neq 0, P_1+P_2= 0)
\lb{eqn:froe-naive-res1}\\
x_j^{n}&=& a_j + 
\ve\bigg(
\frac{A_j}{4\pi}n^2\sin(2\pi a_j)+\frac{C}{4\pi}n^2\sin(2\pi (a_1+a_2))
\bigg)\non\\
&&+{\cal O}(\ve^2),(\mbox{mod} 1)\quad (P_1=0,P_2=0)
\lb{eqn:froe-naive-res2}
\eeqa
and the solutions at $P_1=0,P_2\neq 0$ are calculated as  
\beqa
x_1^n&=&a_1 + 
\ve\bigg(\frac{ A_1}{4\pi}n^2\sin(2\pi a_1)
+\frac{C}{4\pi}\frac{\sin(2\pi a_1+2\pi (a_2+nP_2))}{\cos(2\pi P_2)-1}
\bigg)\non\\
&&+{\cal O}(\ve^2),(\mbox{mod} 1)
\lb{eqn:froe-naive-x1}\\
x_2^n&=&a_2+nP_2
+\ve\bigg(
\frac{A_2}{4\pi}\frac{\sin(2\pi (a_2+nP_2))}{\cos(2\pi P_2)-1}\non\\
&&
+\frac{C}{4\pi}\frac{\sin(2\pi a_1+ 2\pi (a_2+nP_2))}{\cos(2\pi P_2)-1}
\bigg)+{\cal O}(\ve^2),(\mbox{mod} 1).
\lb{eqn:froe-naive-x2}
\eeqa
The solutions at $P_1\neq 0,P_2=0$ 
are obtained by exchanging the suffixes 
$1$ and $2$ in Eqs. \fr{eqn:froe-naive-x1} and \fr{eqn:froe-naive-x2}.
We consider the case that the low-dimensional resonance island appears located 
at $P_1=0,P_2\neq 0$ 
so that we can easily  check the validity of 
our analysis numerically. 
If one considers the case \fr{eqn:froe-naive-nr}, 
then there is no resonance island up to this approximation.
Furthermore one considers the cases \fr{eqn:froe-naive-res1} 
and \fr{eqn:froe-naive-res2}, 
there are resonance islands described by $4$-dimensional maps.
For the case of $P_1=0,P_2\neq 0$, 
a secular term $(\propto \ve n^2)$
appears in Eq.\fr{eqn:froe-naive-x1}, however, does not appear in Eq. 
\fr{eqn:froe-naive-x2}. 
To remove the secular term in Eq.\fr{eqn:froe-naive-x1}, 
we define the renormalized variable 
\beq
a_1^n:=a_1+\ve\frac{A_1}{4\pi}n^2\sin(2\pi a_1).
\lb{eqn:froe-def-a1n}
\eeq 
From the definition \fr{eqn:froe-def-a1n},   
we obtain the symplectic renormalization map
\beq
La_1^n=\ve \frac{A_1}{2\pi}\sin(2\pi a_1^n).
\lb{eqn:froe-rgmap}
\eeq
This procedure to obtain the renormalization map \fr{eqn:froe-rgmap}
is the same as the case 
for the standard map in \S\ref{sec:standard-map}.  
The renormalization map \fr{eqn:froe-rgmap} 
is symplectic and has the form of the standard map.
Since the renormalization map of the standard map is the standard map 
itself (see Eqs.\fr{eqn:stdmp} and \fr{eqn:stdmp-rgmap}), 
a self-similar structure of the phase space could be obtained when one 
considers the renormalization map  \fr{eqn:froe-rgmap}.

From Eqs.\fr{eqn:froe-naive-x1}-\fr{eqn:froe-naive-x2} and 
Eq. \fr{eqn:froe-def-a1n},
we have the following relation 
between the renormalized variable $a_1^n$ 
and the original variables $(x_1^n,x_2^n)$
\beqa
x_1^n&=&a_1^n + 
\ve\frac{ C}{4\pi}\frac{\sin(2\pi a_1^n+2\pi (a_2+nP_2))}{\cos(2\pi P_2)-1}
+{\cal O}(\ve^2),\quad(\mbox{mod} 1)
\lb{eqn:froe-recons-x1}\\
x_2^n&=&a_2+nP_2+\ve
\bigg(
\frac{A_2}{4\pi}\frac{\sin(2\pi (a_2+nP_2))}{\cos(2\pi P_2)-1}
+\frac{C}{4\pi}\frac{\sin(2\pi a_1^n+ 2\pi (a_2+nP_2))}{\cos(2\pi P_2)-1}
\bigg)\non\\
&&\hspace*{2.5cm}+{\cal O}(\ve^2),\quad (\mbox{mod} 1)
\lb{eqn:froe-recons-x2}
\eeqa
%%%%%%%%%%%%%%%
\begin{figure}[h]
\includegraphics[width=6.0cm]
%{/home/sgoto/c-src/Stdmap/Froeschle/P1-0x1x2y2-rg-paper1.eps}
{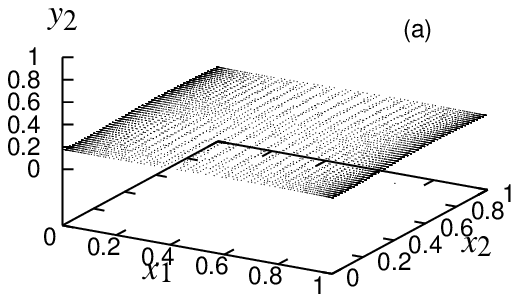}
\includegraphics[width=6.0cm]
%{/home/sgoto/c-src/Stdmap/Froeschle/P1-0x1x2y2-ori-paper1.eps}
{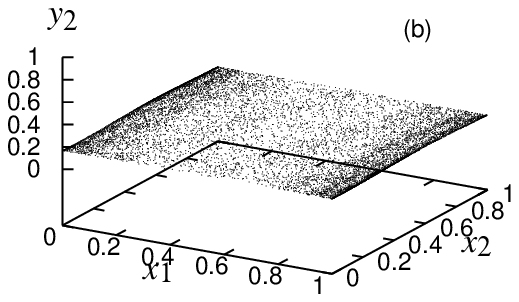}
\caption{
The phase portrait near the resonance island characterized by 
$P_1=0,P_2\neq 0$, 
parameters are
$P_2=0.18(\neq 0), \ve A_1=0.04,\ve A_2=0.05,\ve C=0.02$. 
The number of the initial condition is $1$.
(a): obtained by the renormalization method, and 
(b): calculated using the original map.
}
\lb{fig:froe-x1x2y2}  
\end{figure}
%%%%%%%%%%%%
%%%%%%%%%%%%%%%
\begin{figure}[h]
\includegraphics[width=6.0cm]
%{/home/sgoto/c-src/Stdmap/Froeschle/P1-0x1y1x2-rg-paper1.eps}
{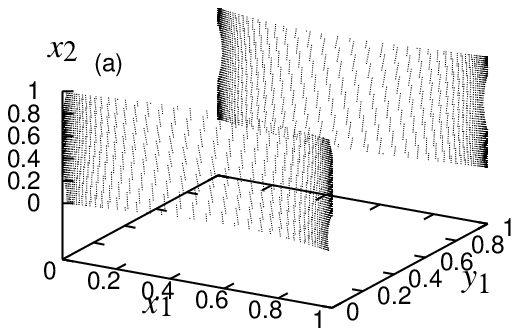}
\includegraphics[width=6.0cm]
%{/home/sgoto/c-src/Stdmap/Froeschle/P1-0x1y1x2-ori-paper1.eps}
{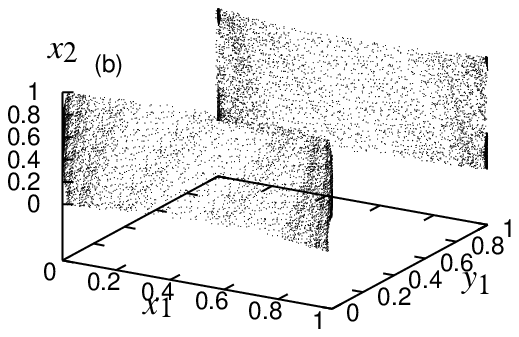}
\caption{
The phase portrait near the resonance island characterized by 
$P_1=0,P_2\neq 0$.
The parameters 
are given in the caption to Fig. \ref{fig:froe-x1x2y2}.
(a): obtained by the renormalization method, and 
(b): calculated using the original map.
}
\lb{fig:froe-x1y1x2}  
\end{figure}
%%%%%%%%%%%%
and the other original variables $y_1^n$ and $y_2^n$ are obtained as 
$y_1^n=x_1^n-x_1^{n-1}$ and $y_2^n=x_2^n-x_2^{n-1}$. 
The unstable manifold of the renormalization map \fr{eqn:froe-rgmap} 
and Eqs. \fr{eqn:froe-recons-x1}--\fr{eqn:froe-recons-x2} reproduce the 
low-dimensional resonance island in the ambient phase space of the system.
Figs. \ref{fig:froe-x1x2y2} and \ref{fig:froe-x1y1x2} show 
that our analysis can predict the low-dimensional resonance 
island even in $4$-dimensional phase space.
Due to this success of the prediction, we understand that 
the low-dimensional resonance island has the same functional form 
as in the $2$-dimensional standard map.  
To confirm the 
%% added below %%
description of the 
%%%%%%%%%%%%%%%%% 
approximate low-dimensional resonant island further, 
we plot a projection of the 
manifold onto the space spanned by $x_1$ and $y_1$ as shown in 
Fig. \ref{fig:froe-x1y1}, 
and it certainly suggests the success of the analysis 
%% %% 11/25 (I wish to add the following sentences)
%% the period in the above sentence should be removed.
 with the perturbation strength $0<|\ve|\lesssim 0.01$.  
 Although we do not show any figure, 
 our perturbative analysis fails
 when $\ve$ is bigger than $\ve_*\sim 0.01$ for $P_1=0$ and $P_2=0.18$.    
 Taking into account Fig. \ref{fig:stdmp}, one finds that  
 the regime $0<|\ve|\lesssim\ve_*$ in which we can use a perturbation analysis 
 for the Froeschl\'e map is narrower than
 that for the standard map in \S\ref{sec:standard-map}.   
 A mathematical estimate is needed to determine the actual validity of our 
 analysis again, as stated in \S\ref{sec:standard-map}.
 On the other hand, 
 according to the expressions for the naive perturbation solutions 
 \fr{eqn:froe-naive-res2}--\fr{eqn:froe-naive-x2},   
  it is natural to expect that the renormalization map \fr{eqn:froe-rgmap}
   fails when $P_2$ approaches to $0$,    
 and we have numerically confirmed that $\ve_*$ generally depends on $P_2$ 
 for $P_1=0$.  
 A situation where the integration constant $P_2$ narrows the regime 
 $0<|\ve|\lesssim\ve_*$ does not appear in $2$-dimensional maps, and 
 is one of features in high-dimensional maps.
%%
%\vspace*{-10mm}

%%%%%%%%%%%%%%%
\begin{figure}[h]
\includegraphics[width=6cm]
%{/home/sgoto/c-src/Stdmap/Froeschle/P1-0x1y1-rg-paper1.eps}
{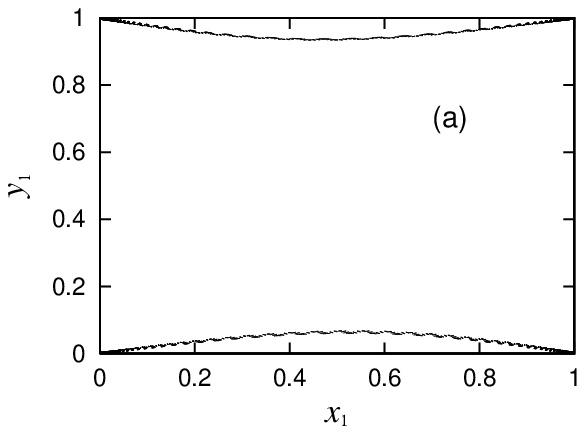}
\includegraphics[width=6cm]
%{/home/sgoto/c-src/Stdmap/Froeschle/P1-0x1y1-org-paper1.eps}
{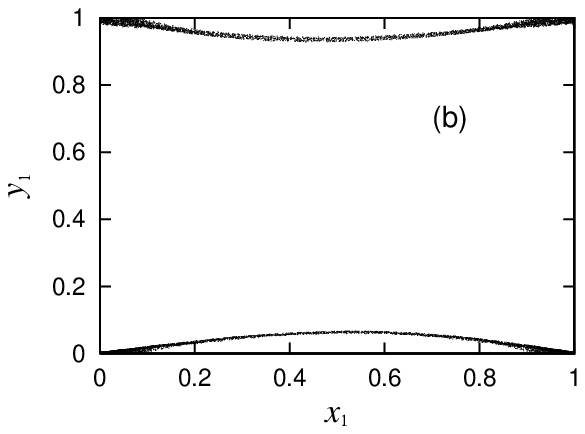}
\caption{
% THIS IS FOR PR
%The phase portrait near the resonance island characterized by 
%$P_1=0,P_2\neq 0$ obtained by 
%Eqs. \fr{eqn:froe-rgmap}-\fr{eqn:froe-recons-x2} and 
%$y_j^n=x_j^n-x_j^{n-1}$,(j=1,2) . 
%The other parameters are given in the caption to Fig. \ref{fig:froe-x1x2y2}.
%%The values of parameters are 
%%$P_2=0.18(\neq 0, \ve A_1 =0.04,\ve A_2=0.05$ and $\ve C=0.02$.
%The inset shows the phase portrait by calculated using the original map.
The phase portrait near the resonance island characterized by 
$P_1=0,P_2\neq 0$.
The parameters 
are given in the caption to Fig. \ref{fig:froe-x1x2y2}.
(a): obtained by the renormalization method, and 
(b): calculated using the original map.
}
\lb{fig:froe-x1y1}  
\end{figure}
%%%%%%%%%%%%

%%%%%%%%%%%%%%%%%%%%%%%%%%%%%%%%%%%%%%%%%%5
\section{Conclusions}
%%%%%%%%%%%%%%%%%%%%%%%%%%%%%%%%%%%%%%%%%%5
 
In this article, 
we have analyzed a  $2$- and a $4$-dimensional symplectic maps using 
a renormalization method. For both maps, we have successfully reduced their
resonance island analytically and the perturbative reductions have been  
confirmed numerically. 
Although  a resonance island is the most fundamental object to study global 
Hamiltonian chaos for a system with any degrees of freedom, 
the island structure cannot be visualized for a system with 
many degrees of freedom. This study shows how a  
low-dimensional island appears in the Froeschl\'e map, 
at least in a nearly integrable regime. 
The method which we have used here is also useful for a wider class of 
symplectic maps, and then this analysis shed light on the study of the 
Hamiltonian systems with many degrees of freedom.

\section*{Acknowledgments}
%We would like to thank ...........

The author thanks the members of the laboratory of dynamical systems theory,
Kyoto University, for their comments and encouragements. 
The author has been supported by
a JSPS Fellowship for Young Scientists.

%\appendix
%\section{First Appendix} %Empty argument \section{} yields `Appendix'. 
%
%\section{Second Appendix}

\end{document}